\documentclass[aps,prr,twocolumn,superscriptaddress,amsmath,amssymb,nofootinbib]{revtex4-1}
\usepackage{graphicx,enumitem}
\usepackage{standalone}
\usepackage{multirow,dcolumn}
\usepackage{txfonts}
\usepackage{hyperref}
\usepackage{soul,cancel}
\usepackage{epsf,amsmath,amssymb,verbatim,color,multirow,pifont,graphicx,cleveref,tabularx}
\usepackage[dvipsnames]{xcolor}
\usepackage{rotating}
\usepackage{amsmath,amsfonts,amssymb,bm,cancel}

\newcommand{\pauliz}[1]{\hat{\sigma}_{#1}^{z}}

\usepackage{tikz}
\usetikzlibrary{matrix}
\usepackage{braket}

\begin{document}
\title{Resolving mean-field solutions of dissipative phase transitions using permutational symmetry} 
\author{Minjae Jo}
\affiliation{QOLS, Blackett Laboratory, Imperial College London SW7 2AZ, UK}
\affiliation{Center for Complex Systems and KI for Grid Modernization, Korea Institute of Energy Technology, Naju, Jeonnam 58217, Korea}

\author{Bukyoung Jhun}
\affiliation{CTP and Department of Physics and Astronomy, Seoul National University, Seoul 08826, Korea}
	
\author{B. Kahng}
\email{bkahng@snu.ac.kr}
\affiliation{Center for Complex Systems and KI for Grid Modernization, Korea Institute of Energy Technology, Naju, Jeonnam 58217, Korea}

	
\begin{abstract}
{Phase transitions in dissipative quantum systems have been investigated using various analytical approaches, particularly in the mean-field (MF) limit. However, analytical results often depend on specific methodologies. For instance, Keldysh formalism shows that the dissipative transverse Ising (DTI) model exhibits a discontinuous transition at the upper critical dimension, $d_c= 3$, whereas the fluctuationless MF approach predicts a continuous transition in infinite dimensions ($d_\infty$). These two solutions cannot be reconciled because the MF solutions above $d_c$ should be identical. This necessitates a numerical verification. However, numerical studies on large systems may not be feasible because of the exponential increase in computational complexity as $\mathcal{O}(2^{2N})$ with system size $N$. Here, we note that because spins can be regarded as being fully connected at $d_\infty$, the spin indices can be permutation invariant, and the number of quantum states can be considerably contracted with the computational complexity  $\mathcal{O}(N^3)$. The Lindblad equation is transformed into a dynamic equation based on the contracted states. Applying the Runge--Kutta algorithm to the dynamic equation, we obtain all the critical exponents, including the dynamic exponent $z\approx 0.5$. Moreover, since the DTI model has $\mathbb{Z}_2$ symmetry, the hyperscaling relation has the form  $2\beta+\gamma=\nu(d+z)$, we obtain the relation $d_c+z=4$ in the MF limit. Hence, $d_c\approx 3.5$; thus, the discontinuous transition at $d=3$ cannot be treated as an MF solution. We conclude that the permutation invariance at $d_\infty$ can be used effectively to check the validity of an analytic MF solution in quantum phase transitions.}
\end{abstract}

	
	
	\maketitle
	
	
\section{Introduction} 
The phase transitions and critical phenomena in dissipative quantum many-body systems have recently attracted considerable attention because theoretical results can be realized experimentally and vice versa ~\cite{carusotto2013,noh2016,carmichael2015,baumann2010,baumann2011,bloch2005,fink2017,fink2018,fitzpatrick2017,espigares2017,helmrich2020,lee2013,jin2016,boite2013,klinder2015,zou2014,nagy2015,houck2012}. 
The mutual competition between the coherent Hamiltonian and incoherent dissipation dynamics creates unexpected emergent phenomena, such as time crystals~\cite{choi2017,gambetta2019}, zero-entropy entangled states~\cite{kraus2008,verstraete2009}, driven-dissipative strong correlations~\cite{tomita2017,ma2019}, and dissipative phase transitions in the nonequilibrium steady state~\cite{sieberer2013,sieberer2014,diehl2010,torre2010,torre2012,tauber2014,sierant2021}, including novel universal behaviors~\cite{marino2016,jo2021}. 
	
Dissipative phase transitions from a disordered (absorbing) state to an ordered (active) state in dissipative quantum systems, such as the quantum contact process (QCP) and the dissipative transverse Ising (DTI) model, have been exploited by developing several analytical techniques in the mean-field (MF) limit. For instance, the Keldysh (or semiclassical MF) approach and fluctuationless MF approach have been proposed. 
In the Keldysh approach, the spins of the DTI model are changed to bosonic operators and an MF functional integral formalism is applied~\cite{sieberer2016, maghrebi2016}. After the upper critical dimension ($d_c$) is determined, a transition point is obtained. In the fluctuationless MF approach, the MF concept is applied to the correlation function. The average product of a pair of individual field amplitudes is treated as the product of the individual averages of the field amplitudes. This result is regarded as a valid approximation in infinite dimensions ($d_\infty$). In addition, noise effects are ignored. In the semiclassical approach, averaging is applied to individuals, as in the fluctuationless MF approach. However, noise effects are considered~\cite{kamenev2011}. These approaches are considered to provide a general framework for exploring the critical behaviors of dissipative phase transitions in the MF limit~\cite{sieberer2016,buchhold2017,sieberer2014}.	 

According to the conventional theoretical framework of equilibrium systems, the two MF solutions at $d_c$ and $d_\infty$ exhibit the same universal behavior. 
For the QCP model, the MF solutions obtained using the semiclassical and fluctuationless MF approaches appear to be the same, as expected. However, for the DTI model, the Keldysh solution predicts $d_c=3$, at which a dissipative phase transition is of the first order when the dissipation is strong, whereas it is of the second order when the dissipation is weak~\cite{maghrebi2016}. In contrast, the fluctuationless MF approach predicts a dissipative phase transition of the second order regardless of the dissipation strength. This result is regarded as the MF solution for $d_\infty$. Accordingly, the two solutions in the strong dissipation limit at $d_c$ and $d_\infty$ are inconsistent. This result was also obtained numerically in three dimensions~\cite{overbeck2017}. Therefore, this discrepancy remains a challenging problem.
	

To resolve this inconsistency, it is necessary to confirm the analytical results numerically. However, numerical approaches, including quantum jump Monte Carlo simulations~\cite{pleino1998}, tensor networks~\cite{vidal2003,verstraete2004a}, and its variants~\cite{verstraete2004b,kshetrimayum2017,werner2016}, are not feasible in higher dimensions because the computational complexity increases exponentially with dimensionality.  
	
Here, {we aim to show that numerical studies are possible when the quantum states can be contracted significantly. Thus, MF solutions for the DTI model can be tested using this numerical method.} For this purpose, we use spin indices that are permutation invariant (PI) on fully connected graphs~\cite{shammah2018}, which are regarded as the graphs at $d_\infty$. {On the all-to-all graphs, the quantum states that are PI can be contracted to a single state. For simplicity, the contracted quantum states are called{\ it PI states}. This contraction considerably reduces the computational complexity} from $\mathcal{O}(2^{2N})$ to $\mathcal{O}(N^3)$, which enables us to numerically study the model in large systems (up to $N=1024$). In this study, we tested the transition type of the DTI model, which was revealed to be continuous. The critical behaviors, which were obtained using finite-size scaling (FSS) analysis, were consistent with those obtained using the fluctuationless MF approach. 

To check the validity of {the numerical method,} we first considered the QCP model~\cite{marcuzzi2016,jo2019,jo2021,carollo2019,gillman2019,gillman2020,gillman2020_2}. This model was chosen because it is regarded as a prototypical model that exhibits dissipative phase transition. Using the semiclassical method and fluctuationless MF approach~\cite{buchhold2017, jo2019}, analytical solutions were obtained at $d_c$ and $d_\infty$. Unlike the DTI model, the two analytical solutions exhibited a continuous transition with the same universal behavior. However, similar to the DTI model, the transition behaviors of the QCP model have not yet been numerically studied because of numerical complexity. Therefore, we performed numerical {studies based on the PI states} and confirmed their agreement with the analytical solutions of $d_c$ and critical exponents.

Next, we consider the transverse Ising (TI) model in a closed quantum system, which corresponds to the zero limit of the dissipation strength of the DTI model in an open quantum system. Because the system is a closed quantum system, we reset the Schr{\"o}dinger equation based on the PI states. We found that its complexity is reduced to $\mathcal{O}(N)$. The static critical exponents obtained were consistent with those reported previously.
		

This study is organized as follows. First, { we introduce the PI states and construct the density matrix based on the PI states in Sec.~\ref{sec:sec2}.} The Lindbald equation of the density matrix is rewritten in the form of the Liouville equation for the PI states. We implement numerical studies for the QCP model using the PI states in Sec.~\ref{sec:qcp}. In Sec.~\ref{sec:qising}, we convert all quantum states of the Schr{\"o}dinger equation to the PI states and implement numerical studies for the TI model. In Sec.~\ref{sec:dti_model}, we perform numerical studies on the DTI model based on PI states and numerically determine the upper critical dimension $d_c$ and static and dynamic critical exponents. In Sec.~\ref{sec:sec5}, we perform numerical simulations using the quantum jump Monte Carlo method for the DTI model with small system sizes and compare the numerical results with those obtained in Sec.~\ref{sec:dti_model}. This additional simulation verifies the numerical method using the PI states. Finally, we present the summary and final remarks in Sec.~\ref{sec:sec6}. 

\begin{table*}
\caption{Summary of previous analytical results. We considered three problems with the quantum models. For each model, the system Hamiltonian ($\hat{H}_S$) and Lindblad operators ($\hat{L}_\ell$) were defined. Semiclassical, Weiss, and Keldysh field-theoretic approaches were used for the QCP, TI, and DTI models, respectively. Among these methods, the Keldysh formalism predicts features that differ qualitatively from those of fluctuationless MF theory.}
\begin{center}
\setlength{\tabcolsep}{12pt}
{\renewcommand{\arraystretch}{1.5}
\vskip 2mm
\begin{tabular}{ccccc}
\hline
\hline
Model & Hamiltonian and Lindblad operators & Field-theoretic approach & fluctuationless MF\\
\hline
\multirow{2}{*}{QCP}  & $\hat{H}_S= \frac{\omega}{N-1} \sum_{m\neq \ell} \hat{n}_m (\hat{\sigma}^+_\ell+\hat{\sigma}^-_\ell) \,, 
$ & Continuous and discontinuous & Continuous and discontinuous  \\ 
& $\hat{L}_{\ell}^{(d)} = \sqrt{\Gamma} \hat{\sigma}_{\ell}^{-} \,, \hat{L}_{m\ell}^{(b)}= \sqrt{\kappa} \hat{n}_m\hat{\sigma}^+_{\ell} \,, \hat{L}_{m\ell}^{(c)}= \sqrt{\kappa} \hat{n}_m\hat{\sigma}^-_{\ell} \,.$ & transitions & transitions \\\hline 
\multirow{2}{*}{TI} 	   & \multirow{2}{*}{$\hat{H}_S=-\frac{J}{N-1}\sum_{\ell \neq m} \hat{\sigma}^z_{\ell}\hat{\sigma}^z_{m}+\Delta\sum_\ell\hat{\sigma}^x_{\ell}\,.$} 
& \multirow{2}{*}{Continuous transition} & \multirow{2}{*}{Continuous transition} \\ 
&  &  & \\ \hline
\multirow{2}{*}{DTI}   & $    \hat{H}_S=-\frac{J}{N-1}\sum_{\ell \neq m} \hat{\sigma}^z_{\ell}\hat{\sigma}^z_{m}+\Delta\sum_\ell\hat{\sigma}^x_{\ell}\,,$ & Discontinuous transition  & \multirow{2}{*}{Continuous transition} \\
& $\hat{L}_{\ell} = \sqrt{\Gamma} \hat{\sigma}_{\ell}^{x^-}\,.$ & with sufficiently strong dissipation &  \\ 
\hline
\hline
\end{tabular}}
\label{tab:tab3}
\label{table}
\end{center}
\end{table*}

\section{Permutational symmetry}\label{sec:sec2}
The time evolution of an open quantum system is described by the Lindblad equation, which comprises the Hamiltonian and dissipation terms:
\begin{align}
\label{eq:lindeq}
\partial_t\hat{\rho}&=-i\left[ \hat{H}_S,\hat{\rho} \right]	+ \sum_{\ell=1}^N\left[ \hat{L}_{\ell}\hat{\rho} \hat{L}^{\dagger}_{\ell}
-\frac{1}{2} \left\{ \hat{L}^{\dagger}_{\ell}\hat{L}_{\ell},\hat{\rho} \right\} \right]\,,
\end{align}
where $\hat{\rho}$, $\hat{H}_S$, and $\hat{L}_\ell$ denote the density matrix of the complete system, system Hamiltonian, and Lindblad operator at the site $\ell$, respectively.

Qubit systems on a fully connected structure are invariant under permutations of the spin indices. The elements of the density matrix satisfy the relation $\rho_{v,w}=\rho_{P(v),P(w)}$, where $v$ and $w$ denote two states among the $2^N$ quantum states of $N$ spins and $P$ denotes a permutation operator. If both the dynamical equation and initial density matrix are PI, the density matrix is also PI.
For example, in a four-spin system, $\rho_{\uparrow \uparrow \downarrow \downarrow, \uparrow \downarrow \uparrow \downarrow} = \rho_{\uparrow \uparrow \downarrow \downarrow, \uparrow \downarrow \downarrow \uparrow} = \rho_{\uparrow \uparrow \downarrow \downarrow, \downarrow \uparrow \downarrow \uparrow}  = \rho_{\downarrow \uparrow \uparrow \downarrow, \uparrow \uparrow \downarrow \downarrow}  = \cdots$. Based on this symmetry, the elements $\left|v\right>\left<w\right|$ of the density matrix can be classified in terms of $(n_1, n_2, s)$, where $n_1$ is the number of up spins in $v$, $n_2$ is the number of up spins in $w$, and $s$ is the number of sites with up spins in both $v$ and $w$ states. Then, the density matrix is written as
\begin{align}
\hat{\rho}=\sum_{n_1, n_2, s}A_{n_1, n_2, s}|n_1,s\rangle\langle n_2,s|\,,
\label{eq:dm}
\end{align}
where $A_{n_1, n_2, s}=\langle n_1,s|\,\hat{\rho}\,|n_2,s\rangle$ is the $3$-rank tensor, whose components are the sum of the elements of $\hat{\rho}$. {$|n_1,s\rangle$ denotes a PI state. In particular, $P(n)$ denotes $A_{n_1=n_2=s=n}$, which represents the probability that the system has $n$ up spins.} For convenience, we introduce a Liouvillian superoperator $\mathcal{L}$ and rewrite the time evolution of the Lindblad equation, Eq.~\eqref{eq:lindeq}, in the form of the Liouville equation: 
\begin{align}
\partial_t\hat{\rho}&=\mathcal{L}\hat{\rho}\,,
\end{align}
This transformation is possible because the Lindblad equation is linear in $\rho$. Consequently, 
\begin{align}
\sum_{n_1,n_2,s}\partial_t A_{n_1, n_2, s}|n_1,s\rangle\langle n_2,s|&=\sum_{n_1,n_2,s}\mathcal{L}A_{n_1, n_2, s} |n_1,s\rangle\langle n_2,s|\,.
\label{eq:tensoreq}
\end{align}
Thus, the computational complexity decreases as $\mathcal{O}(N^3)$~\cite{shammah2018}.

\section{Quantum contact process}
\label{sec:qcp}

We consider the QCP model~\cite{marcuzzi2016,jo2019,jo2021,carollo2019,gillman2019,gillman2020,gillman2020_2}, which is a paradigmatic model exhibiting an absorbing phase transition in open quantum systems. This theoretical model has recently attracted the attention of scientists because it is simple and can thus be analytically solved at $d_c$ and $d_\infty$. Moreover, it has been realized experimentally in ultracold Rydberg atomic systems using the antiblockade effect~\cite{gutierrez2017} in the classical limit. However, the numerical results of this model in the MF limit have not yet been obtained because of its numerical complexity. 
We performed numerical studies using the Runge-Kutta algorithm for the Liouville equation~\eqref{eq:tensoreq} based on the PI states.  

The Hamiltonian $\hat{H}_S$ contains coherent terms for branching and coagulation and is given by
\begin{align}
\hat{H}_S= \frac{\omega}{N-1} \sum_{m\neq \ell} \hat{n}_m (\hat{\sigma}^+_\ell+\hat{\sigma}^-_\ell) \,.
\label{eq:HS_QCP}
\end{align}
The Lindblad decay, branching, and coagulation operators are given by
\begin{align}
\label{eq:ld_QCP}
\hat{L}_{m\ell}^{(b)}= \sqrt{\kappa} \hat{n}_m\hat{\sigma}^+_{\ell} \,, \quad
\hat{L}_{m\ell}^{(c)}= \sqrt{\kappa} \hat{n}_m\hat{\sigma}^-_{\ell} \,, \quad
\hat{L}_{\ell}^{(d)} = \sqrt{\Gamma} \hat{\sigma}_{\ell}^{-} \,, 
\end{align}
respectively.
{Here, $\hat{n}_\ell= \left| \uparrow \rangle \langle \uparrow \right|_\ell$ is the number operator of the active state at site $\ell$ and $\hat{\sigma}^{\pm}_\ell=(\hat{\sigma}^x_\ell\pm i\hat{\sigma}^y_\ell)/2$.}
The composite operator $\hat{n}_m\hat{\sigma}^+_{\ell}$ or $\hat{n}_m\hat{\sigma}^-_{\ell}$ with $\ell \neq m$ indicates that the active state at site $m$ activates or deactivates the state at site $\ell$, which represents the branching or coagulation processes, respectively. $\kappa$ is the rate of incoherent branching or coagulation. In contrast, $\hat{L}_{\ell}^{(d)}$ in Eq.~\eqref{eq:ld_QCP} denotes the decay dynamics $\ket{\uparrow} \to \ket{\downarrow}$ at $\ell$, where $\Gamma$ is the decay rate. Therefore, if there is no active state, no further dynamics occur and the system enters an absorbing state.
	
\begin{figure*}
\includegraphics[width=0.8\linewidth]{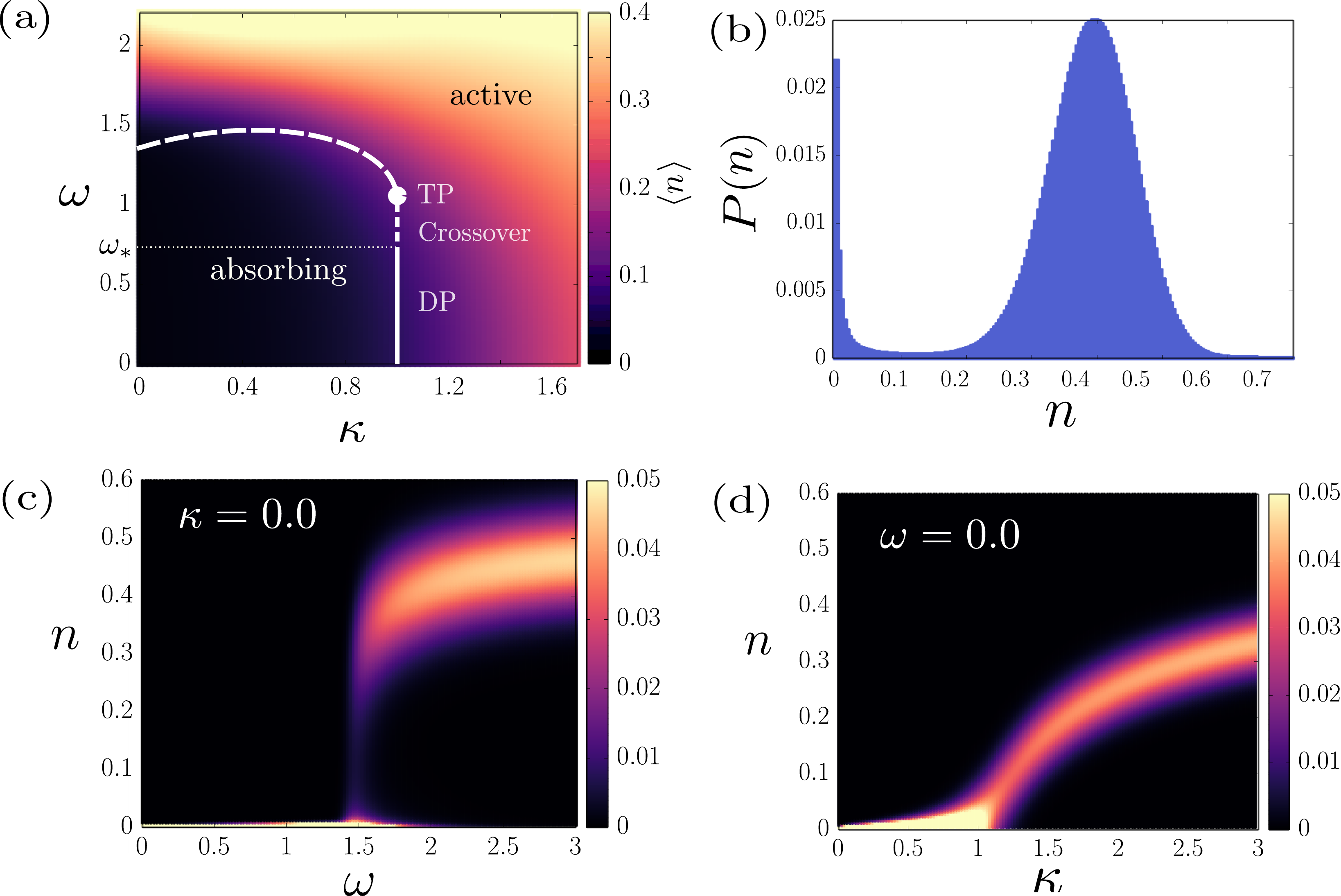}
\caption{(a) Phase diagram of the QCP model on a fully connected graph in the parameter space ($\kappa$, $\omega$), {determined by the direct numerical enumeration of the Liouville equation based on the PI states}. Discontinuous (dashed curve) and continuous transitions (dotted and solid lines) occur, and they meet at a TP. {On the dotted critical line in $\omega \in [\omega_*,1]$, the critical exponent $\alpha$ varies continuously, whereas on the solid line, it has a DP value. The color indicates the average order parameter $\langle n \rangle$ defined in Eq.~\eqref{average_n}).} (b) Distribution of the order parameters $n$ for $(\kappa,\omega)=(0,1.8)$. This indicates that the system is bistable at $n=0$ and $\approx 0.45$. (c) Plot of the order parameter $n$ as a function of $\omega$ for $\kappa=0.0$ in the steady state. This shows that the transition is of the first order.  (d) Order parameter $n$ plotted as a function of $\kappa$ at $\omega=0.0$ in the steady state. This shows that the transition is of the second order. Data were obtained by direct enumeration from a system of size $N=256$. We set $\Gamma=1$ for all the figures. In (c) and (d), the color indicates the probability that the order parameter $n$ exists.}
\label{fig:fig1}
\end{figure*}

According to the MF solution obtained using the semiclassical method~\cite{buchhold2017,jo2019}, the QCP exhibits three types of phase transitions: i) for $\kappa < 1$, a discontinuous transition [dashed line in Fig.~\ref{fig:fig1}(a)] occurs; ii) for $\kappa_c=1$ and $\omega_* < \omega < \omega_c=1$ [dotted line in Fig.~\ref{fig:fig1}(a)], a continuous transition occurs with continuously varying exponents; and iii) for $\kappa=1$ and $\omega < \omega_*$, a continuous transition [solid line in Fig.~\ref{fig:fig1}(a)] occurs. A tricritical point (TP) appears at $(\kappa_c, \omega_c)$, as shown in Fig.~\ref{fig:fig1}. The continuous transition iii) belongs to the directed percolation (DP) universality class~\cite{cardy1980}. The continuous transition at the TP belongs to the tricritical DP class~\cite{grassberger2006,lubeck2006,jo2020}. 
	
\begin{figure}[!t]
\includegraphics[width=1.0\columnwidth]{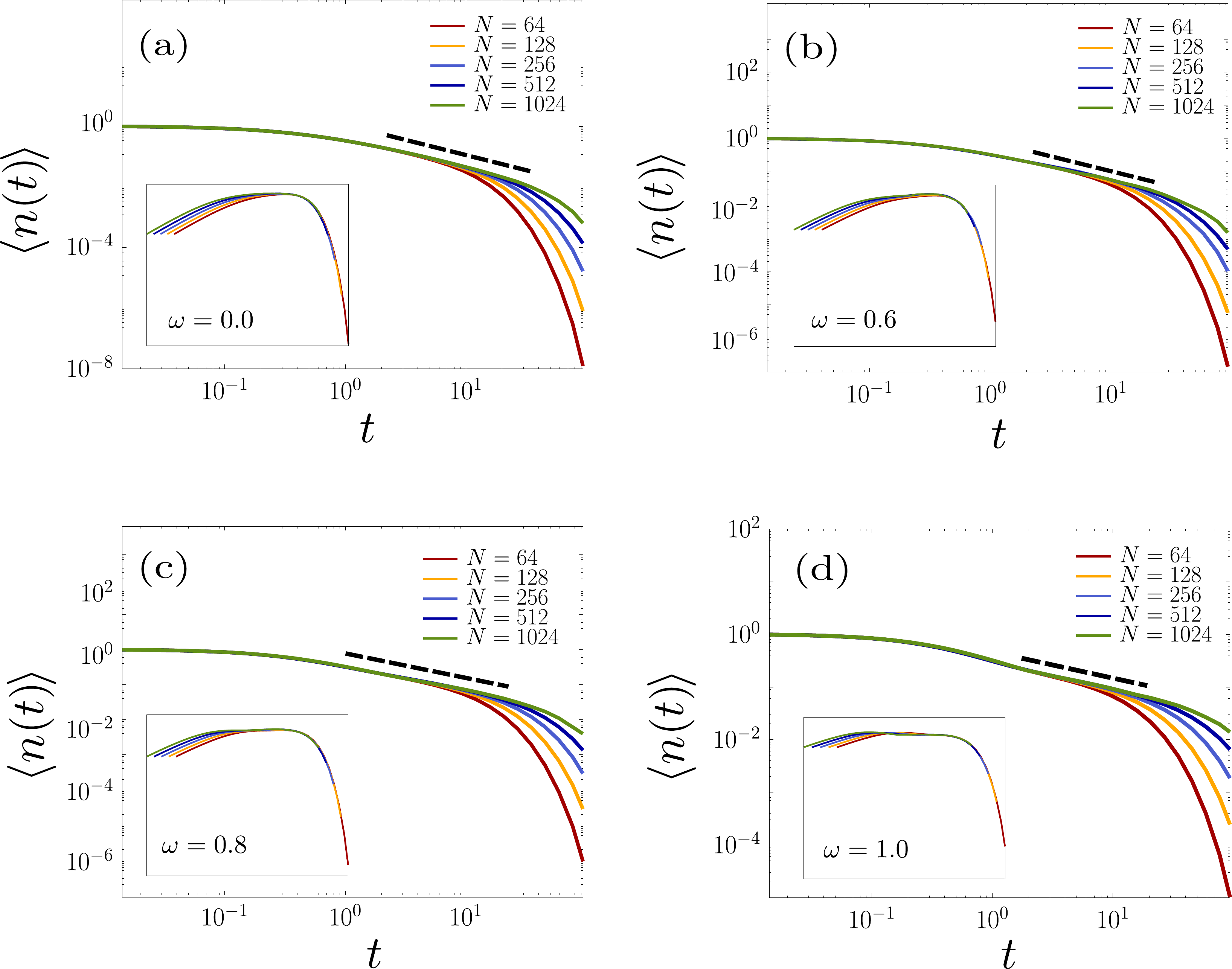}
\caption{Plots of the order parameter $\langle n(t)\rangle$ of the QCP model as a function of time $t$ for fixed $\kappa=1$ but different (a) $\omega=$0.0, (b) 0.6, (c) 0.8, and (d) 1.0. Direct numerical enumeration of the Liouville equation based on the PI states was employed. As $N$ increases, the data exhibits power-law behavior in the large-$t$ regime. The guide lines (dashed lines) show the power-law behavior, indicating that $\langle n(t)\rangle \sim t^{-\alpha}$. The exponent $\alpha$ was estimated as (a) $\alpha=1.0$ for $\omega=0$, (b)  $\alpha=0.92$ for $\omega=0.6$, (c) $\alpha=0.70$ for $\omega=0.8$, and (d) $\alpha=0.50$ for $\omega=1.0$. Figures (a)--(d) show that the critical exponent $\alpha$ varies continuously, depending on $\omega$. Data were obtained for $\Gamma=1$ and $\kappa=1$. Insets: Scaling plots of $\langle n(t)\rangle t^{\alpha}$ versus $tN^{-z}$.
\label{fig:fig2}}
\end{figure}

\begin{table}[!h]
\begin{center}
\caption{Critical exponent $\alpha$ values for different {$\omega$} values.}
\vskip 2mm
\setlength{\tabcolsep}{25pt}
{\renewcommand{\arraystretch}{1.3}
\begin{tabular}{cc}
\hline
\hline
$\omega$ & $\alpha$ \\
\hline
$1.0$ & $0.50\pm 0.02$  \\
$0.9$ & $0.61\pm 0.02$ \\
$0.8$ & $0.70\pm 0.02$ \\
$0.7$ & $0.81\pm 0.02$ \\
$0.6$ & $0.92\pm 0.02$ \\
$\le 0.53$ &  {MF DP values}\\
\hline
\hline
\end{tabular}}
\label{tab:tab1}
\end{center}
\end{table}
	
	
We discuss the numerical results of the QCP model based on PI states. The QCP model exhibited a phase transition from the absorbing state to the active state, as shown in Fig.~\ref{fig:fig1}]. The order parameter of the phase transition is defined as the average density of active sites (i.e., the sites of up spins), formulated as   
\begin{align} 
\langle n(t) \rangle =\Big(\sum_{\ell}\text{Tr}[\hat{\rho}(t)\hat{n}_{\ell}]\Big)/N=\sum_n P(n) n \,.
\label{average_n}
\end{align} 
In the absorbing state, $\langle n(t) \rangle \to 0$ as $t\to \infty$, whereas in the active state, $\langle n(t) \rangle \to$ is finite as $t\to \infty$. The phase boundaries comprised two parts for the first- and second-order transitions in the parameter space $[\kappa, \omega]$, and their positions were consistent with those predicted by the theory using the semiclassical method. 

{The numerical method using the PI states enables the easy calculation of $P(n)$ as a function of $n$ for any given $\kappa$ and $\omega$, as shown in Fig.~\ref{fig:fig1}(b). The density of $n$ up spins is broadly distributed around the phase boundary. The two stable stationary solutions at $n=0$ and $n\approx 0.45$ indicate a first-order transition.}     

The numerical results were obtained using a fully connected graph of size $N=256$. Along the continuous transition line (solid line) at $\kappa=1$, as shown in Fig.~\ref{fig:fig1}(a), we examined the critical behavior under different initial conditions. For an initial state comprising all up spins at time $t=0$, we measured $\langle n(t) \rangle$ as a function of time for different system sizes up to $N=1024$. We find that $\langle n(t)\rangle$ exhibits a power-law decay as $\langle n(t)\rangle\sim t^{-\alpha}$. As predicted by the theory, the exponent $\alpha$ varies continuously for $\omega_* < \omega < 1$ with $\omega_*\approx 0.53$, as shown in Fig.~\ref{fig:fig2}(d)$-$(b)]. $\alpha$ is fixed at $1.0\pm 0.02$ at $\omega=0.0$ [Fig.~\ref{fig:fig2}(a)] and $\alpha=1$ is the DP value. Numerical estimates for different $\omega$ values are listed in Table~\ref{tab:tab1}). Therefore, we conclude that the numerical method based on the PI states successfully reproduces the theoretical values of the QCP model.

\section{Dissipative Transverse Ising model}
\label{sec:sec4}

\subsection{Transverse Ising model}
\label{sec:qising}
The Hamiltonian $\hat{H_S}$ of the TI model at $d_\infty$ is expressed as
	\begin{align}
	\hat{H}_S=-\frac{J}{N-1}\sum_{\ell \neq m} \hat{\sigma}^z_{\ell}\hat{\sigma}^z_{m}+\Delta\sum_\ell\hat{\sigma}^x_{\ell}\,,
	\label{eq:def_hamiltonian}
	\end{align}
where $J$ represents the strength of the ferromagnetic interaction of the Ising spins in the $z$ direction. The summation runs for every pair of spins. $\ell$ is the spin index, $\ell=1,\cdots N$. The parameter $\Delta$ represents the strength of the transverse field. 
When $\Delta/J < 1$, the ferromagnetic interaction becomes dominant and the ground states are two-fold degenerate ordered states, whereas for $\Delta/J > 1$, the ground state is nondegenerate and disordered. Thus, the system exhibits a quantum phase transition~\cite{sachdev2011} from a ferromagnetic ($\Delta/J < 1$) to a paramagnetic phase ($\Delta/J > 1$). 
It is to be noted that this Hamiltonian has $\mathbb{Z}_2$ symmetry under the transformation $\hat{\sigma}^z \to -\hat{\sigma}^z$.
	
The Liouville equation (Eq.~\eqref{eq:tensoreq}) must be replaced by imaginary-time dynamics, because the TI model is a closed quantum system. The elements of the wavefunction satisfy the relation $\psi_\nu = \psi_{P(\nu)}$, where $\nu$ denotes a state among the $2^N$ quantum states of $N$ spins and $P$ denotes a permutation operator. 
Therefore, the wave function is simply written as
\begin{equation}
\ket{\psi} = \sum_{n=0}^N B_n\ket{n} \,,
\end{equation}
where $n$ is the number of up spins in $\nu$ and $B_n$ is the coefficient of state $n$. Thus, we must track only $N+1$ complex numbers to study the system.
	
To obtain the ground state, we used the imaginary-time Schr\"odinger evolution $\partial_t \ket{\psi} = -\hat{H_S} \ket{\psi}$ under the normalization condition for the wave function $\braket{\psi | \psi} = 1$. Using the above expression for the wave function, we obtain the following differential equations for $B_n$:
\begin{equation}
\sum_{n=0}^N \partial_t B_n \ket{n} =- \sum_{n=0}^N B_n {\hat{H}_S \ket{n}} \,. \label{eq:diff_eq_closed}
\end{equation}
Unlike the Lindblad open quantum systems, where the normalization condition $\sum_\nu \rho_{\nu \nu} = \sum_n A_{n,n,n} = 1$ holds owing to the dynamics given by Eq.~\eqref{eq:tensoreq}, the normalization condition $\sum_{n=0}^N \left|B_n\right|^2 = 1$ is broken at each time step. Therefore, $B_n$ must be rescaled at each time step in the simulation to restore normalization.
	
\begin{figure}
\includegraphics[width=0.8\linewidth]{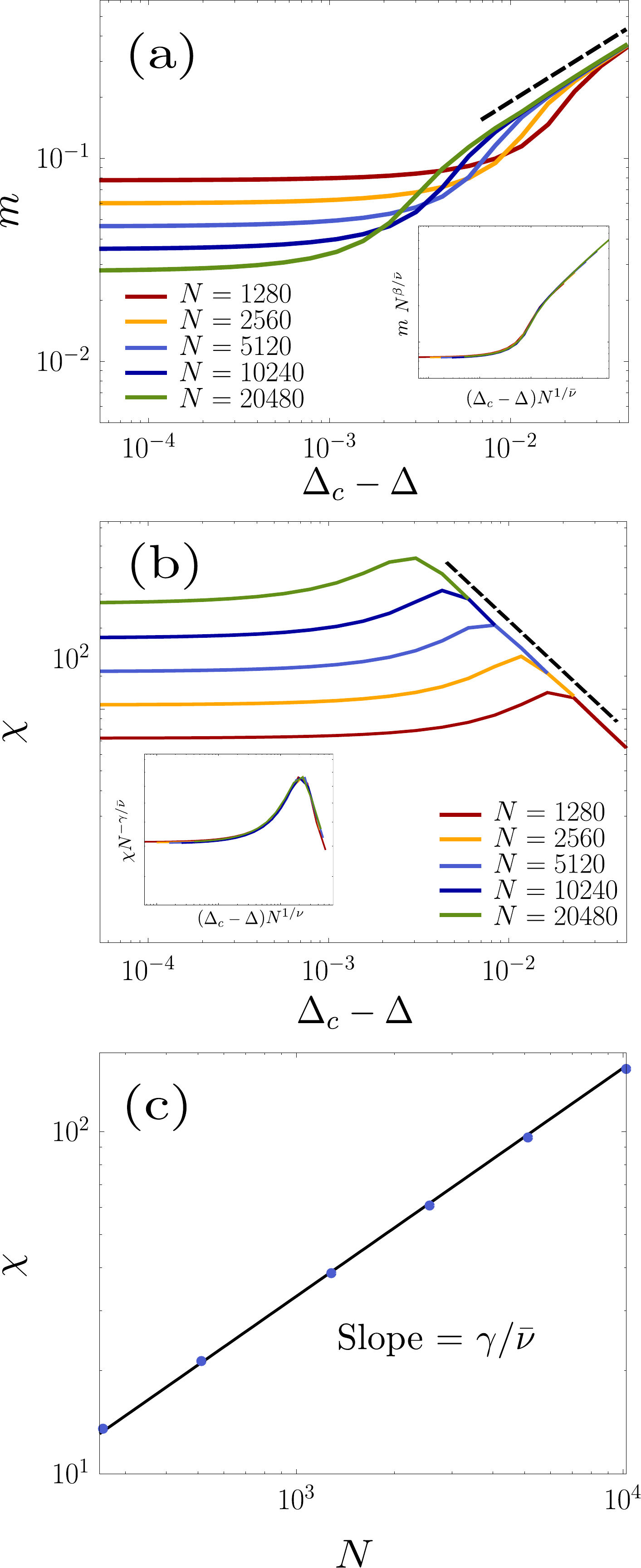}	
\caption{FSS analysis for the transverse Ising model on fully connected graphs. Direct numerical enumeration of the Liouville equation based on the PI states was employed. (a) Order parameter $m$ plotted as a function of $\Delta_c-\Delta$ for different system sizes. We set $J=1$ and $\Delta_c = 1$.
The auxiliary dashed line with a slope of $0.5$ indicates power-law behavior $m \sim \left(\Delta_c-\Delta\right)^{0.5}$. Inset: Scaling plot of magnetization $m N^{\beta/\bar{\nu}}$ versus $\left(\Delta_c-\Delta\right) N^{1/\bar{\nu}}$, with $\bar{\nu}=1.5$ and $\beta=0.5$. 
(b) Plot of the susceptibility as a function of $\Delta_c-\Delta$ for different system sizes. $\chi = N^{1+\bar{z}} \left( \left< \left(\hat{\sigma}^z\right)^2 \right> - \left< \hat{\sigma}^z \right>^2 \right)$, where $\bar{z}=z/d_c=1/3$. {The black dashed line is a guide line indicating $\chi \sim \left(\Delta_c-\Delta\right)^{-1}$.} Inset: Scaling plot of susceptibility $\chi N^{-\gamma/\bar{\nu}}$ versus $\left(\Delta_c-\Delta\right) N^{1/\bar{\nu}}$.
The critical exponents $\bar{\nu}=1.5$ and $\gamma=1.0$ were used for the FSS analysis.
(c) Plot of $\chi$ versus $N$ at $\Delta_c$. The slope represents the value of the critical exponent $\gamma/{\bar \nu}$.
}
\label{fig:fig3}
\end{figure}

Using this method, we perform numerical iterations of the dynamics, as shown in Eq.~\eqref{eq:diff_eq_closed}, for different system sizes. FSS analysis measures the critical exponents $\beta$ and $\bar \nu$ associated with the order parameter and correlation size, respectively. For a steady state of $B_n$, the magnetization is obtained as
\begin{align}
m=\sum_n |B_n|^2|m_n|\,,
\label{eq:def_magnetization}
\end{align}
where $m_n \equiv (1/N)\bra{n}\sum_i \pauliz{i}\ket{n}$. 
We plot the magnetization $m$ versus $\Delta_c-\Delta$ for different sizes of $N$ up to $N=20480$ in Fig.~\ref{fig:fig3}(a) and obtain the critical exponent $\beta=0.50\pm 0.01$. We also plot $mN^{\beta/\bar{\nu}}$ versus $\left(\Delta_c-\Delta\right)N^{1/\bar{\nu}}$ in the inset of Fig.~\ref{fig:fig3}(a). In this plot, $\bar \nu$ is considered such that the data points for various $N$ values collapse onto a single curve. $\bar{\nu}=d_c\nu=1.5\pm 0.01$ is obtained. 

The susceptibility $\chi$~\cite{pang2019}, which represents the fluctuations of the order parameter in finite quantum systems, is defined as 
\begin{align}
\chi=N^{1+\bar{z}}\left(\langle m^2\rangle-\langle m\rangle^2 \right)\,,
\label{eq:def_susceptibility}
\end{align}
where $z=\bar z d_c$ represents the dynamical critical exponent. We note that the dynamical exponent contributes to the susceptibility because the imaginary time appears as an extra dimension at zero temperature, and the dynamic correlation function appears with the imaginary time axis~\cite{boite2013} in a closed quantum system. Thus, critical phenomena are described using an additional scaling variable with a single new exponent $z$~\cite{tauber2014_book}.
The susceptibility diverges as $\chi\sim (\Delta_c-\Delta)^{-\gamma}$ as $\Delta \to  \Delta_c^-$. Therefore, we plot $\chi$ versus $\Delta_c-\Delta$ on a double logarithmic scale and find that $\chi$ exhibits a power-law decay with a slope of $-1.00\pm 0.01$. Thus, the exponent $\gamma$ is estimated as $\gamma=1.00\pm 0.01$. In particular, it should be noted that in Fig.~\ref{fig:fig3}(b), the data points collapse onto a single power-law line in the subcritical region when ${\bar z}=0.33\pm 0.005$. Inserting this $\bar{z}$ value into $d_c+z=4$ (see Appendix~\ref{appnedixB} for more details), we obtain $d_c=3.0\pm 0.01$ and $z=1.0 \pm 0.03$. At $\Delta=\Delta_c$, $\chi \sim N^{\gamma/\bar \nu}$ holds, as shown in Fig.~\ref{fig:fig3}(c). A similar plot was presented in Ref.~\cite{pang2019} for the one-dimensional case with $L=120$. Next, plotting $\chi N^{-\gamma/\bar{\nu}}$ versus $(\Delta_c-\Delta)N^{1/\bar{\nu}}$ and taking $\bar{\nu}=1.50\pm 0.01$, we find that the data points collapse onto a single curve. These results confirm that $\bar \nu=1.50\pm 0.01$. When $\bar{z}$ is chosen as the classical Ising value, $\bar{z}=0$ in Eq.~\eqref{eq:def_susceptibility}, we confirm that the data collapse fails owing to the incorrect value of $\bar{z}$.

Next, the dimensional analysis of Eq.~\eqref{eq:def_susceptibility}, we find the hyperscaling relation $2\beta+\gamma=\nu(d_c+z)$, or equivalently, $2\beta+\gamma=\bar \nu(1+\bar z)$~\cite{dutta2015}. {In Appendix~\ref{appnedixB}, we derive the hyperscaling relation explicitly from the correlation function.} Using the numerical values of $\beta=0.50\pm 0.01$, $\gamma=1.00\pm 0.04$, $\bar{\nu}=1.50\pm 0.01$, and $\bar z=0.33\pm 0.01$, we find that the hyperscaling relation is satisfied. Below, we examine whether this hyperscaling relationship remains valid in dissipative quantum systems.

\subsection{Dissipative transverse Ising model}
\label{sec:dti_model}

\subsubsection{Model definition}

We considered the DTI model~\cite{ates2012,jin2018,rose2016,hu2013}, which has been experimentally realized using ultracold Rydberg atoms~\cite{malossi2014,carr2013}.
For open quantum systems, in addition to the Hamiltonian $\hat{H}_S$, given by Eq.~\eqref{eq:def_hamiltonian}, 
 the Lindblad operator must account for the dissipation process. For the DTI model, spin decay was imposed from positive to negative eigenvectors on the $x$ axis. This operation can be expressed as follows:
\begin{align} 
\hat{L}_{\ell} = \sqrt{\Gamma} \hat{\sigma}_{\ell}^{x^-}=\sqrt{\Gamma} \frac{\hat{\sigma}_{\ell}^{z}+i\hat{\sigma}_{\ell}^{y}}{2}, 
\label{eq:Lindblad_ope}
\end{align} 
where $\Gamma$ is the decay rate. 
 This system retains the $\mathbb{Z}_2$ symmetry  under the transformation  $(\hat{\sigma}^x,\hat{\sigma}^y,\hat{\sigma}^z)\to(\hat{\sigma}^x,-\hat{\sigma}^y,-\hat{\sigma}^z)$. Accordingly, the critical exponents of static variables are expected to remain in the Ising class~\cite{tauber2014_book}.
Conventionally, this DTI model is known to exhibit a continuous transition according to the fluctuationless MF approach~\cite{ates2012}. However, a recent analytical solution based on the Keldysh formalism shows that the transition at the upper critical dimension (which is calculated as $d_c=3$) is not continuous but rather discontinuous when the dissipation is sufficiently strong~\cite{maghrebi2016}, specifically, in the regions $\Delta/\Gamma<0.5$~\cite{maghrebi2016} and $\Delta/J< 0.22$~\cite{overbeck2017}. This result was confirmed by numerical results obtained using the variational method~\cite{overbeck2017}.

\subsubsection{Fluctuationless MF approach}
\label{sec:dti_mf}
Let us consider the MF solution for the DTI model using the fluctuationless MF approach~\cite{jo2019,buchhold2017}.
To obtain the MF solution, the equation of motion for an observable $ O$ can be explored. This equation is given by the conjugate master equation:
\begin{align}
\partial_t\hat{O}&=i\left[ \hat{H}_S,\hat{O} \right]
+ \sum_{\ell=1}^N\left[ \hat{L}^{\dagger}_{\ell}\hat{O} \hat{L}_{\ell}
-\frac{1}{2} \left\{ \hat{L}^{\dagger}_{\ell}\hat{L}_{\ell},\hat{O} \right\} \right]\,.
\end{align}
Ignoring correlations and assuming uniform fields, we derive the MF equations as follows: 
\begin{align}
\label{eq:mf_x}
\partial_t{\sigma}^x &= 4J\sigma^y\sigma^z-\Gamma(1+\sigma^x) \,, \\
\label{eq:mf_y}
\partial_t{\sigma}^y &= -4J\sigma^x\sigma^z -2\Delta\sigma^z-\frac{\Gamma}{2}\sigma^y\,, \\
\label{eq:mf_z}
\partial_t{\sigma}^z &= 2\Delta \sigma^y-\frac{\Gamma}{2}\sigma^z \,.
\end{align}
We find that two sets of steady-state solutions exist for $\{\sigma^x_0, \sigma_0^y, \sigma^z_0\}$. The first set is given as: 
\begin{align}
\label{eq:sol1}
\sigma^x_0=-1\,,\quad\sigma^y_0=\sigma^z_0=0\,,
\end{align}
and the other set is given as 
\begin{align}
\label{eq:sol2}
\sigma^x_0 &= \frac{J}{\Delta}(\sigma^z_0)^2-1\,, \qquad
\sigma^y_0=\frac{4}{\Gamma}\left(2J-\Delta\right)\left(1+\frac{32J^2z^2}{\Gamma^2}\right)^{-1}\sigma^z_0\,, \nonumber\\
\sigma^z_0 &= \pm \frac{\sqrt{-\Gamma^2+32J\Delta-16\Delta^2}}{4\sqrt{2}J} \,. 
\end{align}
	
Subsequently, we check the stability of the two solutions. For the first solution  [Eq.~\eqref{eq:sol1}], the dynamical equations~\eqref{eq:mf_x}--\eqref{eq:mf_z} are linearized around the fixed point. Inserting $\sigma^x = \sigma^x_0 + \delta \sigma^x$, $\sigma^y = \sigma^y_0 + \delta \sigma^y$, and $\sigma^z = \sigma^z_0 + \delta \sigma^z$ into Eqs.~\eqref{eq:mf_x}--\eqref{eq:mf_z}, and expanding up to the linear order of perturbations, we obtain the linear equation $\dot{\delta{\bf a}} = {\bf M} \delta{\bf a}$, where 
\begin{equation}\label{eq:linear}
\delta{\bf a} = \left( \delta \sigma^x_0, \delta \sigma^y, \delta \sigma^z \right)^{\sf T}\,,
\end{equation}
and the matrix ${\bf M}$ is given by
\begin{equation}
\label{eq:matrix}
{\bf M}=\left(\begin{array}{ccc} -\Gamma & 4J\sigma^z_0 & 4J\sigma^y_0\\ 
-4J \sigma^z_0 & -\frac{\Gamma}{2} & -2\Delta-4J\sigma^x_0\\ 
0 & 2\Delta&  -\frac{\Gamma}{2}
\end{array}\right).
\end{equation}
All the eigenvalues of ${\bf M}$ are negative in the region $2J-\Delta-\Gamma^2/(16\Delta)<0$, indicating a stable fixed point. 

For the other solution [Eq.~\eqref{eq:sol2}], all the eigenvalues of ${\bf M}$ are negative in the other region, $2J-\Delta-\Gamma^2/(16\Delta)>0$, indicating an unstable fixed point. Thus, a continuous phase transition occurs from the disordered phase governed by Eq.~\eqref{eq:sol1} to the ordered phase governed by Eq.~\eqref{eq:sol2} across the transition line given by
\begin{align} 
\label{eq:transition_line}
2J-\Delta-\Gamma^2/(16\Delta)=0\,.
\end{align}
	
Substituting the expression for $\sigma^y_0$ in Eq.~\eqref{eq:sol2} into Eq.~\eqref{eq:mf_z}, the equation can be expanded with respect to $\sigma^z_0\ll 1$ as follows:
\begin{align}
\partial_t \sigma^z&=0 = -u_2\sigma^z_0-u_4(\sigma^z_0)^3 +\mathcal{O}\left((\sigma^z_0)^5\right)\,,
\label{eq:SS}
\end{align}
where $u_2$ and $u_4$ are defined as
\begin{align}
u_2 = -\frac{8\Delta}{\Gamma}\left(2J-\Delta-\frac{\Gamma^2}{16\Delta}\right)\,\quad  {\rm and} \quad
u_4=\frac{256\Delta J^2}{\Gamma^3}\left(2J-\Delta\right)\,.
\end{align}
It is to be noted that Eq.~\eqref{eq:SS} implies the existence of an effective potential defined as
\begin{align}
U(\sigma^z)=\frac{u_2}{2}(\sigma^z)^2+\frac{u_4}{4}(\sigma^z)^4+\mathcal{O}\left((\sigma^z)^6\right)\,.
\label{eq:effective_potential}
\end{align}
Here $\mathcal{O}\left((\sigma^z)^6\right)$ is irrelevant because the $u_4$ term is always positive near the transition line. Thus, we consider terms up to the $u_4$ term. It is to be noted that $U(\sigma^z)=U(-\sigma^z)$ holds because of $\mathbb{Z}_2$ symmetry, and this effective potential describes the universality class of the classical Ising model.
Then, the solution $\sigma^z_0$ that satisfies Eq.~\eqref{eq:SS} is also the steady-state solution of the single effective equation of the order parameter, which is given by
\begin{align*}
\partial_t \sigma^z=-\frac{\partial U}{\partial \sigma^z}\,.
\end{align*}
Again, we obtain the transition line of Eq.~\eqref{eq:transition_line} given by $u_2=0$.
Then, the transition line $\Gamma_c$ is expressed as a function of $J$: 
\begin{align}
\Gamma_c/J=4\sqrt{2(\Delta_c/J)-(\Delta_c/J)^2}\,,
\label{eq:phase_boundary_J_1}
\end{align}
where $\Delta_c$ denotes the value of $\Delta$ at the transition line.
The order parameter behavior near the transition line can then be obtained by considering the minimum value of the effective potential in Eq.~\eqref{eq:effective_potential}; the resulting solution is $\sigma^z_0$ in Eq.~\eqref{eq:sol2}. Expanding the order parameter for $\Gamma_c-\Gamma \ll 1$, we obtain
\begin{align}
m=\frac{\sqrt{\Gamma_c}}{4J}\sqrt{\Gamma_c-\Gamma}\,,
\label{eq:solution_J_1}
\end{align}
which yields the exponent of magnetization $\beta=0.5$.
Similarly, we find the transition line for fixed $\Gamma$ as follows:
\begin{align}
J_c/\Gamma=\frac{1+16(\Delta_c/\Gamma)^2}{32(\Delta_c/\Gamma)}\,.
\label{eq:phase_boundary_Gamma_1}
\end{align}
We also obtain the order parameter for $J-J_c \ll 1$ as 
\begin{align}
m=\frac{\sqrt{\Delta_c}}{J_c}\sqrt{J-J_c}\,.
\label{eq:solution_Gamma_1}
\end{align}
Thus, the critical exponent $\beta=0.5$ is obtained.

\begin{figure}
\includegraphics[width=0.90\columnwidth]{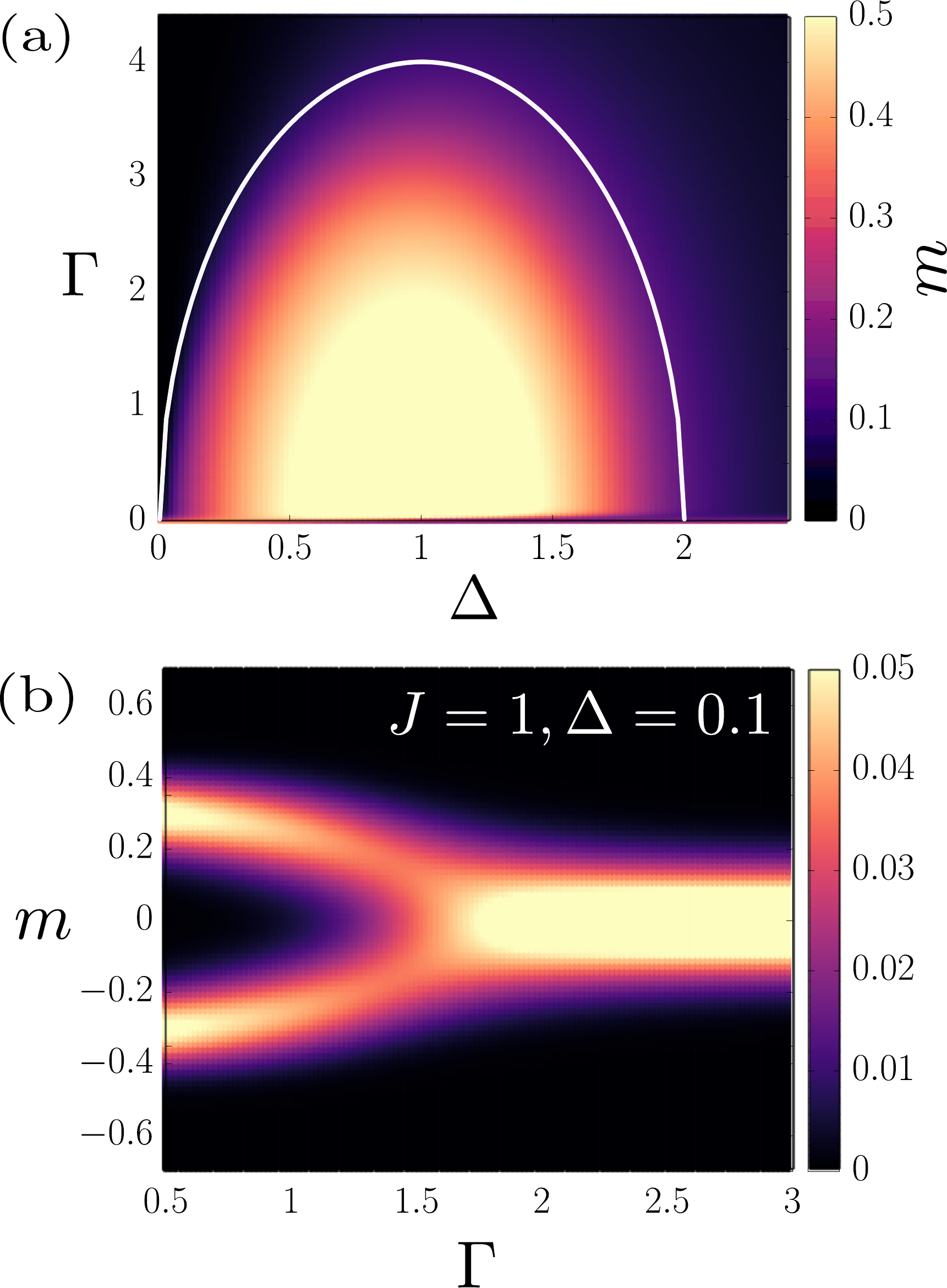}
\caption{(a) Phase diagram of the DTI model on fully connected graphs in the parameter space ($\Delta$, $\Gamma$). A continuous transition occurs across the solid white curve. The brightness represents the magnitude of the magnetization $m$. (b) Density of the order parameter in the steady state as a function of $\Gamma$ at $\Delta=0.1$. The system size was $N=128$. The brightness represents the probability that the order parameter exists. The data were obtained from the Liouville equation with $J=1$ based on the PI states.}
\label{fig:fig4}
\end{figure}
	
\begin{figure}
\includegraphics[width=0.8\columnwidth]{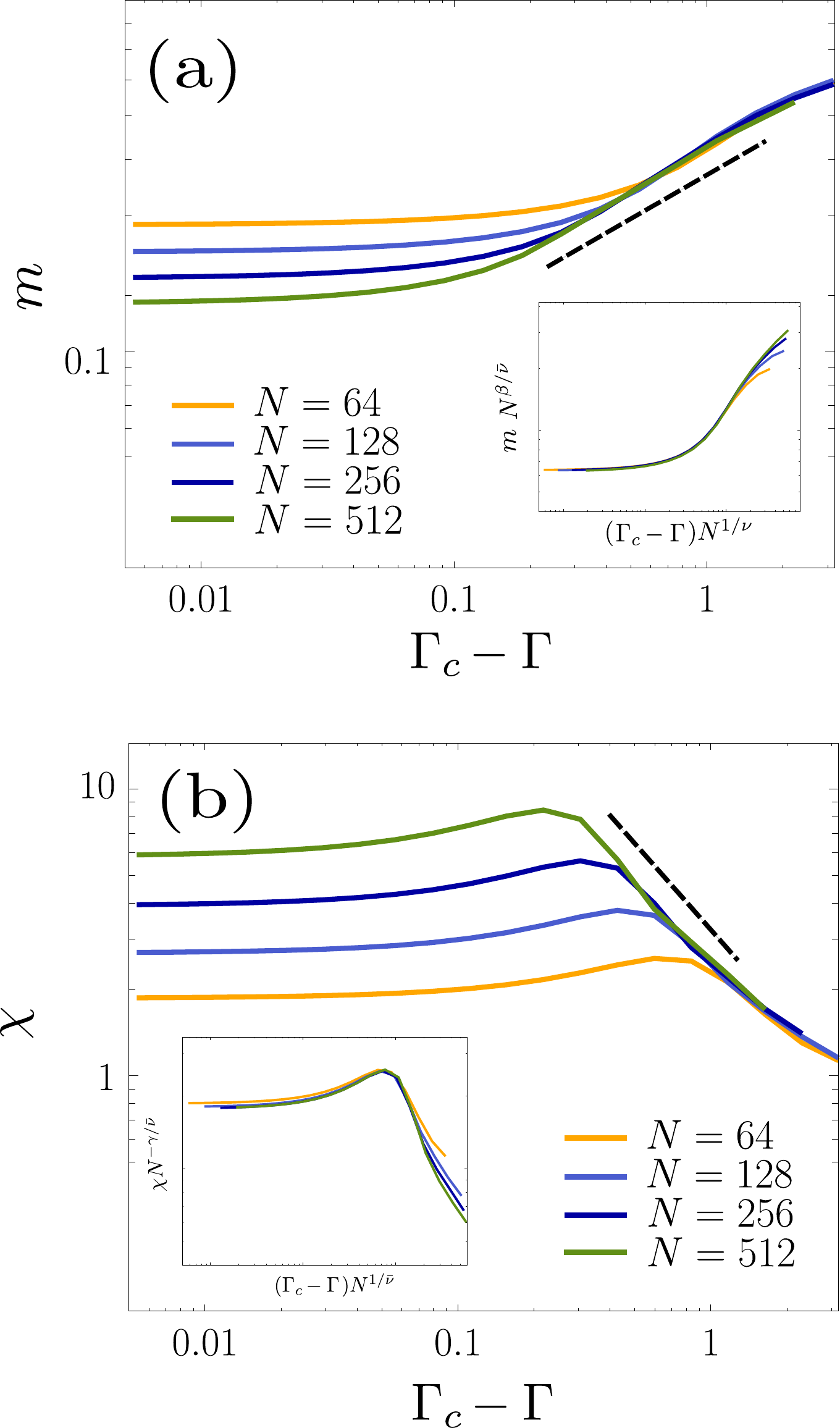}
\caption{FSS analysis for the DTI model at $\Delta=0.5$ and $J=1$ on fully connected graphs. Direct numerical enumeration of the Liouville equation based on the PI states was employed.
(a) Plot of the order parameter $m$ as a function of $\Gamma_c-\Gamma$ for different system sizes.
The auxiliary dashed line with a slope of $0.5$ indicates power-law behavior $m \sim \left( \Gamma_c -\Gamma \right)^{0.5}$. Inset: Scaling plot of the magnetization $m N^{\beta/\bar{\nu}}$ versus $\left(\Delta_c-\Delta\right) N^{1/\bar{\nu}}$ with $\bar{\nu}=1.75$ and $\beta=0.5$. 
(b) Plot of $\chi$ as a function of $\Gamma_c-\Gamma$ for different system sizes. The auxiliary dashed line with a slope of $-1.0$ indicates power-law behavior $\chi\sim (\Gamma_c-\Gamma)^{-1.0}$. Inset: Scaling plot of susceptibility $\chi N^{-\gamma/\bar{\nu}}$ versus $\left(\Gamma_c-\Gamma\right) N^{1/\bar{\nu}}$.
The critical exponents $\bar{\nu}=1.75$ and $\gamma=1.0$ were used for the FSS analysis.
}
\label{fig:fig5}
\end{figure}
	
\subsubsection{Numerical results}
\label{sec:dti_numerical}

\begin{figure}
\includegraphics[width=0.90\columnwidth]{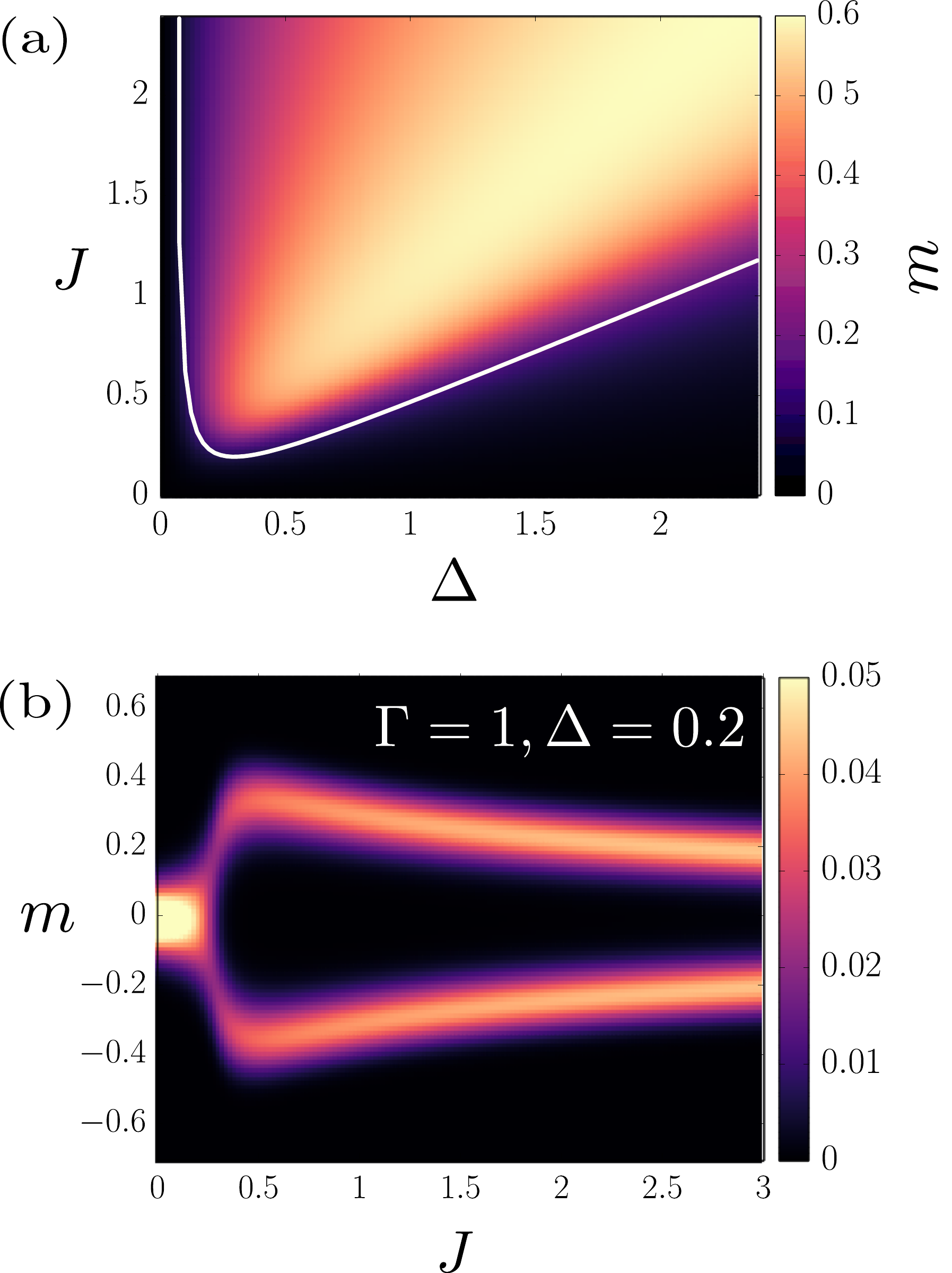}
\caption{
(a) Phase diagram of the DTI model on fully connected graphs in the parameter space ($\Delta$, $J$). A continuous transition occurs across the solid white curve. The brightness represents the magnitude of magnetization $m$.  
(b) Density of the order parameter in steady state as a function of $J$ at $\Delta=0.2$. The system size was $N=128$. The brightness represents the probability that the order parameter exists. The data were obtained from the Liouville equation with $\Gamma=1$ based on the PI states.
}
\label{fig:fig6}
\end{figure}
	
\begin{figure}
\includegraphics[width=0.8\columnwidth]{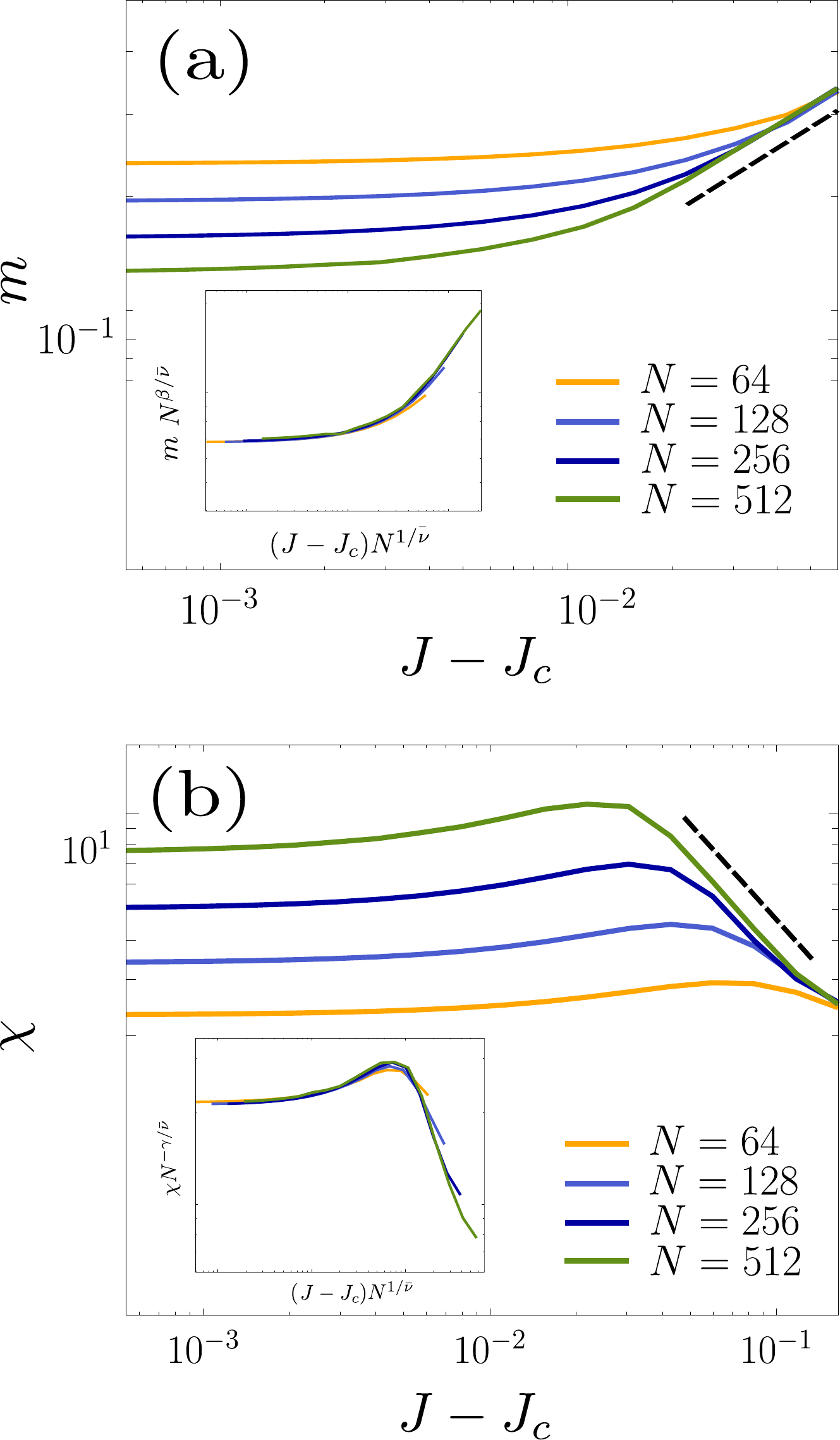}
\caption{FSS analysis for the DTI model at $\Delta=0.2$ and $J=1$ on fully connected graphs. Direct numerical enumeration of the Liouville equation based on the PI states was employed.
(a) Plot of the order parameter $m$ as a function of $J-J_c$ for different system sizes.
The auxiliary dashed line with a slope of $0.5$ indicates power-law behavior $m \sim \left( J-J_c \right)^{0.5}$. Inset: Scaling plot of magnetization $m N^{\beta/\bar{\nu}}$ versus $\left( J-J_c \right) N^{1/\bar{\nu}}$ with $\bar{\nu}=1.75$ and $\beta=0.5$. 
(b) Plot of $\chi$ as a function of $J-J_c$ for different system sizes. The auxiliary dashed line with a slope of $-1.0$ indicates power-law behavior $\chi\sim (J-J_c)^{-1.0}$. Inset: Scaling plot of the susceptibility $\chi N^{-\gamma/\bar{\nu}}$ versus $\left(J-J_c \right) N^{1/\bar{\nu}}$. The critical exponents $\bar{\nu}=1.75$ and $\gamma=1.0$ were used.
}
\label{fig:fig7}
\end{figure}

Hereafter, we consider the numerical results for the DTI model.
We first consider the case wherein $J$ is fixed at $J=1$ but $\Delta$ and $\Gamma$ are varied.
Numerical simulations were performed by applying the Runge-Kutta method to the Liouville equation [Eq.~\eqref{eq:tensoreq}] based on PI states. The phase diagram obtained numerically in parameter space ($\Delta, \Gamma$) is shown in Fig.~\ref{fig:fig4}(a). The phase boundary curve (white curve) was obtained using the analytical fluctuationless MF solution [Eq.~\eqref{eq:phase_boundary_J_1}]. A distribution of the order parameter is shown in Fig.~\ref{fig:fig4}(b), where $J=1$ and $\Delta=0.1$ were fixed.
	
Next, we perform an FSS analysis to obtain the critical exponents $\beta$ and $\bar \nu$, which are associated with the order parameter and correlation size, respectively. 
We obtained the critical exponent $\beta$ associated with the order parameter by directly measuring the local slope of the plot of $m$ versus $\Gamma_c-\Gamma$ on a double logarithmic scale, as shown in Fig.~\ref{fig:fig5}(a). We then plot $mN^{\beta/\bar{\nu}}$ versus $(\Gamma_c-\Gamma)N^{1/\bar{\nu}}$ for different system sizes $N$, as shown in the inset of Fig.~\ref{fig:fig5}(a). This result confirms that $\beta=0.50\pm 0.01$. We also obtained a correlation size exponent $\bar{\nu}$ of $\bar{\nu}=d_c\nu=1.75\pm 0.01$ using the FSS analysis, as shown in the inset of Fig.~\ref{fig:fig5}(a). It is noteworthy that the critical exponent $\beta$ agrees with the analytical result of Eq.~\eqref{eq:solution_J_1}. 

The susceptibility $\chi$ defined in Eq.~\eqref{eq:def_susceptibility} exhibits divergent behavior $\chi\sim (\Gamma_c-\Gamma)^{-\gamma}$ as $\Gamma \to \Gamma_c^-$. 
Therefore, we plot $\chi$ versus $\Delta_c-\Delta$ on a double logarithmic scale and find that $\chi$ exhibits a power-law decay with a slope of $-1.00\pm 0.01$ in Fig.~\ref{fig:fig5}(b). Thus, exponent $\gamma$ is estimated as $\gamma=1.00\pm 0.01$. We note that the data points collapse onto a single power-law line in the subcritical region when ${\bar z}=0.14\pm 0.01$. By inserting this $\bar{z}$ value into $d_c+z=4$ (see  Appendix~\ref{appnedixB} for more details), we obtain $d_c=3.5\pm 0.02$ and $z=0.5\pm 0.03$. Next, we plot $\chi N^{-\gamma/\bar{\nu}}$ versus $(\Delta_c-\Delta)N^{1/\bar{\nu}}$, taking $\bar{\nu}=1.75\pm 0.01$, as shown in the inset of  Fig.~\ref{fig:fig5}(b). We observe that the data points collapse onto a single curve. These results confirm that $\bar \nu=1.75\pm 0.01$. 

Next, we consider the case where $\Gamma=1$ is fixed.
The phase diagram obtained numerically in parameter space ($\Delta, J$) is shown in Fig.~\ref{fig:fig6}(a). The heat map data are obtained by the direct enumeration of the magnetization on the basis of PI states, whereas the phase boundary curve (white curve) is obtained by the fluctuationless MF solution. The probability of the order parameter is shown in Fig.~\ref{fig:fig6}(b), where $\Delta=0.2$ is fixed but $J$ varies. For $\Delta=0.2$, a discontinuous transition was predicted by Keldysh formalism; however, we obtained a continuous transition. The order parameter curve does not increase monotonously; instead it decreases after a point near $J\approx 0.5$. It is likely that the order parameter saturates at a constant value in the large $J$ limit.
	
Next, we perform an FSS analysis to obtain the critical exponents $\beta$ and $\bar \nu$, which are associated with the order parameter and correlation size, respectively. From Fig.~\ref{fig:fig7}(a), we obtain $ \beta = 0.50 \pm 0.01 $ for $\Delta = 0.2$  by measuring the local slope of $m$ as a function of $J-J_c$ on a double logarithmic scale. Here, $J_c$ is the value predicted by the fluctuationless MF theory. Next, we plotted $mN^{\beta/\bar{\nu}}$ versus $(J-J_c)N^{1/\bar{\nu}}$ for different system sizes, as shown in the inset of Fig.~\ref{fig:fig7}(a). In this plot, $\bar \nu$ is the value at which the data for different values of $ N$ collapse onto the same curve. We obtain $\bar{\nu}=d_c\nu=1.75\pm 0.01$. The susceptibility, as defined in Eq.~\eqref{eq:def_susceptibility}, diverges as $\chi\sim (\Delta_c-\Delta)^{-\gamma}$ as $\Delta \to  \Delta_c^-$. Therefore, we plot $\chi$ versus $\Delta_c-\Delta$ on a double logarithmic scale and find that $\chi$ exhibits a power-law decay with a slope of $-1.00\pm 0.01$ in Fig.~\ref{fig:fig7}(b). The data points collapse onto a single power-law line in the subcritical region when ${\bar z}=0.14\pm 0.01$. Next, by plotting $\chi N^{-\gamma/\bar{\nu}}$ versus $(\Delta_c-\Delta)N^{1/\bar{\nu}}$ and considering $\bar{\nu}=1.75\pm 0.01$, we find that the data collapses onto a single curve, as shown in the inset of Fig.~\ref{fig:fig7}(b). This result confirms that $\bar \nu=1.75\pm 0.01$. 

Similarly, we obtain the same critical exponents and upper critical dimension along the transition line. The obtained exponents $\beta=0.50\pm 0.01$, $\gamma=1.00\pm 0.04$, $\bar{\nu}=1.75\pm 0.01$, and $\bar z=0.14\pm 0.01$ satisfy the hyperscaling relation $2\beta+\gamma=\nu(d_c+z)$ or equivalently, $2\beta+\gamma=\bar \nu(1+\bar z)$~\cite{dutta2015}. The Lindblad operator in Eq.~\eqref{eq:Lindblad_ope} conserves $\mathbb{Z}_2$ symmetry; thus, the static critical exponents are the same, whereas it affects the dynamics and the related critical exponent $z$.

\section{Comparison with quantum jump Monte Carlo simulation}
\label{sec:sec5}
\begin{figure}[!t]
\includegraphics[width=0.85\columnwidth]{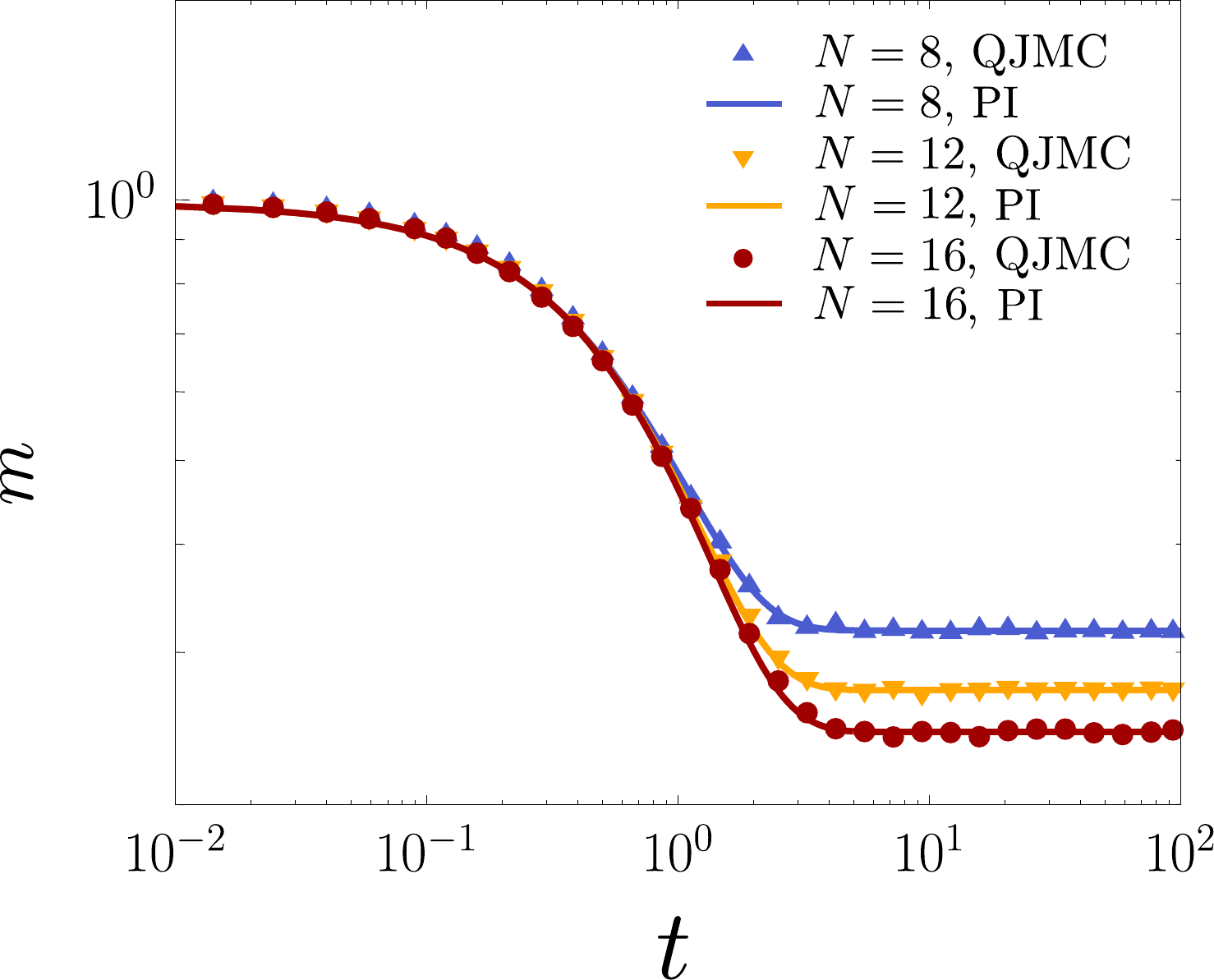}
\caption{Comparison of the data sets obtained by the direct enumerations of the Liouville equation based on the PI state (solid curve) and quantum jump Monte Carlo (QJMC) simulations (symbols) for the DTI model with $\Delta=0.1$ and $J=1$ for different system sizes. 
\label{fig:fig8}}
\end{figure}
	
To verify the validity of the numerical method of the Lindblad equation based on the PI states, we performed quantum jump Monte Carlo simulations for the DTI model on a fully connected graph. Simulations were performed on relatively small system sizes of $N=8$, $12$, and $16$, as shown in Fig.~\ref{fig:fig8}.  The two methods were found to be consistent.
	
\section{Summary and Discussion}
\label{sec:sec6}
\begin{table*}
\caption{Summary of universality classes for classical, closed quantum, and open quantum Ising models. The system Hamiltonian ($\hat{H}_S$) and Lindblad operators ($\hat{L}_\ell$) were defined for each model. Our numerical results indicate that dissipation changes the upper critical dimension and dynamic critical exponent. Note that the hyperscaling relation $2\beta+\gamma=\nu (d_c+z)$ is valid for all the models.}
\begin{center}
\setlength{\tabcolsep}{12pt}
{\renewcommand{\arraystretch}{1.5}
\begin{tabular}{c |c| c| c }
\hline
\hline
Model & Classical Ising model & TI model & DTI model\\ \hline
Hamiltonian&  $\hat{H}=-J\sum_{\langle\ell m\rangle} \hat{\sigma}^z_{\ell}\hat{\sigma}^z_{m}$ & $\hat{H}=-J\sum_{\langle\ell m\rangle} \hat{\sigma}^z_{\ell}\hat{\sigma}^z_{m}+\Delta\sum_\ell\hat{\sigma}^x_{\ell}$ &  				$\hat{H}=-J\sum_{\langle\ell m\rangle} \hat{\sigma}^z_{\ell}\hat{\sigma}^z_{m}+\Delta\sum_\ell\hat{\sigma}^x_{\ell}$ \\ \hline
Lindblad operator&$-$ & $-$ & {$\hat{L}_{\ell} = \sqrt{\Gamma} \hat{\sigma}_{\ell}^{x^-}$}\\
\hline
$d_c$ & 4 & 3 & 3.5  \\
\hline
$z$ & 0 & 1 & 0.5 \\
\hline
Other critical exponents & \multicolumn{3}{l}{~~~~~~~~~~~~~~~~~~~~~~~~~~~~~~~~~~~~~~{$\beta=0.5$, $\gamma=1.0$, $\nu=0.5$, and $d_c+z=4$.}} \\
\hline
\hline
\end{tabular}}
\end{center}
\label{table}
\end{table*}

We performed numerical simulations of the qubit systems at $d_\infty$, where different theoretical methods sometimes yielded inconsistent predictions. In particular, it is conjectured that the phase transitions in the upper critical dimension and infinite dimensions are different~\cite{maghrebi2016}. Considering the equilibrium case, this result is unexpected because phase transitions above the upper critical dimension correspond to the infinite-dimensional case. Using FSS analysis, which has not been feasible because of the complexity of the numerical approach to infinite-dimensional systems, we investigated the critical MF behavior of the DTI model. In addition, we verified {PI-based method} by comparing the time-dynamic behavior obtained from quantum jump Monte Carlo simulations for small qubit sizes (Fig.~\ref{fig:fig8}).
	
We first considered the QCP, where a previous result based on the semiclassical MF solution showed that the continuous transition belongs to the DP universality class and the TP belongs to the tricritical DP class (Fig.~\ref{fig:fig1}).
Using our approach, we found that the transition lines are exactly the same as those obtained using the semiclassical approach with an upper critical dimension $d_c=3$~\cite{jo2019}. Furthermore, there exists a crossover region along which the exponent $\alpha$ (which is associated with the density of active sites) decreases continuously from the tricritical DP value to the DP value, which is reminiscent of one-dimensional QCP~\cite{jo2021}. 
	
Next, both the TI and DTI models are characterized by $\mathbb{Z}_2$ symmetry. Thus, the universality class in the steady state should belong to the Ising universality class with $\beta=0.5$, $\gamma=1.0$, and $\nu=0.5$. We successfully performed FSS analysis using this analytical transition line obtained from the fluctuationless MF results. The critical exponents $\beta\approx 0.5$, $\gamma\approx 1.0$, $\bar \nu\approx 1.5$, and $\bar z\approx 0.33$ were obtained for the TI model. Thus, the upper critical dimension and dynamic critical exponent were determined to be $d_c=3$ and $z=1$, respectively, which are important for quantum phase transitions because the upper critical dimension is smaller by $z$ than that of the classical transition~\cite{boite2013}.
In contrast, for the DTI model, the critical exponents $\beta\approx 0.5$, $\gamma\approx 1.0$, $\bar \nu\approx 1.75$, and $\bar z\approx 0.14$ were obtained. Inserting these values into $d_c+z=4$, we obtained $d_c=3.5\pm 0.02$ and $z=0.5 \pm 0.03$. Thus, both models satisfy the hyperscaling relation $2\beta+\gamma=\nu(d_c+z)$ or equivalently, $2\beta+\gamma=\bar \nu(1+\bar z)$~\cite{dutta2015}. The MF universality behaviors of the three Ising-type models are summarized in Table~\ref{table}.

{Our result implies that if the DTI model is simulated at $d=4$, the transition would be continuous with the criticality in the MF limit. These results differ from those obtained from the Keldysh formalism \mbox{~\cite{maghrebi2016}}. Although the Keldysh field theory is well-justified for bosonic systems, it is limited by bosonization when applied to spin systems. When the Keldysh formalism is applied to spin systems, such as the DTI model, it is necessary to map spins to bosons, for instance, through hard-core bosonization using a large on-site potential, which might yield a valid qubit system in the infinite potential limit.}

In conclusion, we exploited permutation invariance (PI) at $d_\infty$ assuming that at $d_\infty$, spins interact in an all-to-all manner. Owing to the PI, the quantum states contract significantly, which considerably reduced the computational complexity to $\mathcal{O}(N^3)$. Therefore, we performed numerical studies and observed that Keldysh formalism is invalid for the DTI model. {We believe that the PI property can be used for other problems, such as quantum synchronization, arising in all-to-all networks~\cite{mendoza2014}.  }

\begin{acknowledgments}
This research was supported by the NRF (Grant No.~NRF-2014R1A3A2069005), a KENTECH Research Grant (KRG2021-01-007) (BK), and the quantum computing technology development program of the NRF funded by the Ministry of Science and ICT (No.~2021M3H3A103657312) (MJ). M.J. and B.J. contributed equally to this work.
\end{acknowledgments}

	\begin{appendix}
\section{Fluctuationless MF approach for QCP}
\label{appendixA}
One can derive the fluctuationless MF equations for $\sigma^x(t)$, $\sigma^y(t)$, and $n(t)$:
\begin{align}
\dot{n} &= \omega n \sigma^y
+(\kappa -1) n -2\kappa n^2 \,, \nonumber\\
\dot{\sigma}^x &= -\omega \sigma^x\sigma^y
-\frac{1+\kappa }{2}\sigma^x -\kappa n\sigma^x \,,\nonumber \\
\dot{\sigma}^y &= \omega \left\{ 2n
+\left(\sigma^x\right)^2- 4n^2 \right\}
-\frac{1+\kappa}{2}\sigma^y-\kappa n\sigma^y \,,
\end{align}
where we rescale time as $t\Gamma \to t$, $\omega /\Gamma \to \omega$, and
$\kappa/\Gamma \to \kappa$. 
Then, two solutions can be obtained for each region. The first solution becomes
\begin{equation} \label{eq:critical}
\kappa=1 \,{\rm and}\, \omega \le 1 \,,
\end{equation}
and the second solution is
\begin{equation} \label{eq:firstorder}
\omega = \left( 1+\kappa -\kappa^2 + \sqrt{(1+\kappa -\kappa^2 )^2 -\kappa^4 } \right)^{1/2}
\,\,\, {\rm at } \,\,\,
\kappa \le 1\,.
\end{equation}
The first (second) solution is shown as the solid (dashed) line in Fig.~\ref{fig:fig1}(a) in the main text.
		
		

\section{Derivation of hyperscaling relation}
\label{appnedixB}
The scaling ansatz of the free energy density for classical phase transitions is easily adopted for quantum phase transitions. At zero temperature, the (imaginary) time acts as an additional dimension because the extension of the system in this direction is infinite. Therefore, the scaling ansatz of the free energy density at zero temperature is expressed as ~\cite{vojta2003}
\begin{align}
f(g,h)=b^{-(d+z)}f(b^{y_g}g,b^{y_h}h)\,,
\label{eq:1}
\end{align}
where {$b$ is a scale factor}, $g \equiv \Delta_c-\Delta$ is the rescaled transverse field, $h$ is the magnetic field, and $y_g$ and $y_h$ are the scaling exponents.
A comparison of this relation with the classical homogeneity law shows that a quantum phase transition in $d$ spatial dimensions is equivalent to a classical transition in $d + z$ spatial dimensions. Thus, for a quantum phase transition, the upper critical dimension, above which the MF critical behavior becomes exact, is smaller by $z$ than the corresponding classical transition.

To determine the critical exponent $\beta$ of the magnetization, which is defined as $m\propto g^{\beta}$, we differentiate Eq.~\eqref{eq:1} with respect to $h$ and set $h=0$, that is: 
\begin{align}
m\propto \frac{\partial f}{\partial h}\bigg|_{h=0} \propto b^{-(d+z)+y_h}f(b^{y_g}g,0)\,.
\end{align}
By setting $b^{y_g}g=1$, we obtain $m\propto g^{\frac{d+z-y_h}{y_g}}$,  and thus $\beta=(d+z-y_h)/{y_g}$.

Next, to determine the critical exponent of the susceptibility $\chi\propto g^{-\gamma}$, we differentiate Eq.~\eqref{eq:1} twice with respect to $h$ and set $h=0$, that is: 
\begin{align}
\chi\propto \frac{\partial^2 f}{\partial h^2}\bigg|_{h=0} \propto b^{-(d+z)+2y_h}f(b^{y_g}g,0)\,.
\end{align}
By setting $b^{y_g}g=1$, we obtain $\chi\propto g^{\frac{d+z-2y_h}{y_g}}$, and thus $\gamma=(-d-z+2y_h)/y_g$.

Consider the behavior of the correlation length $\xi\propto g^{-\nu}$ under a renormalization group transformation
\begin{align}
\xi\propto b \xi(b^{y_g}g,b^dN^{-1})\,.
\end{align}
When $b^{y_g}g=1$ is chosen and $N\to \infty$, $\xi\propto g^{-\frac{1}{y_g}}$, implying that $\nu={1}/{y_g}$.

Thus, the hyperscaling relation $2\beta+\gamma=\nu(d+z)$, or equivalently $2\beta+\gamma=\bar{\nu}(1+\bar{z})$, holds, where $\bar{\nu}=d\nu$, and $\bar{z}=z/d$.

Next, we consider the dissipative transverse (quantum) Ising (DTI) model. Generally, the dissipator of DTI models can break $\mathbb{Z}_2$ symmetry under the transformation $(\hat \sigma^x, \hat \sigma^y, \hat \sigma^z)\to (\hat \sigma^x, -\hat \sigma^y, -\hat \sigma^z)$. However, the DTI model we consider has $\mathbb{Z}_2$ symmetry, like the classical Ising and the transverse Ising model. Hence, the static critical exponents remain the same as those of the classical Ising model. $(2\beta+\gamma)/\nu=4$, and thus, $z+d_c=4$ in the mean-field limit. 


\end{appendix}



%

\end{document}